\documentstyle[11pt,newpasp,twoside]{article}
\def\edcomment#1{\iffalse\marginpar{\raggedright\sl#1\/}\else\relax\fi}
\reversemarginpar

\newcommand{\Bx}{{\mathbf{x}}} 
\newcommand{\Bv}{{\mathbf{v}}}
\newcommand{\cf}{\bar{f}}
\newcommand{\cv}{\bar{v}}

\def\ifm#1{\relax\ifmmode#1\else$\mathsurround=0pt #1$\fi}
\def\kms{\ifmmode\,{\rm km}\,{\rm s}^{-1}\else km$\,$s$^{-1}$\fi} 
\def\hmpc{\,\ifm{h^{-1}}{\rm Mpc}}

\def\kpc{\,{\rm kpc}}
\def\msun{M_{\odot}}
\def\etal{et al.\ }

\def\Rvir{R_{\rm vir}}
\def\Nvir{M_{\rm vir}}
\def\fvir{f_{\rm vir}}
\def\f02{f_{20\%}}
\def\fdel{f_{\rm del}}
\def\vdel{v_{\rm del}}
\def\Vdel{V_{\rm del}}

\def\ftrue{f_{\rm true}}
\def\vtrue{v_{\rm true}}

\def\pmb#1{\setbox0=\hbox{#1}
 \kern-.025em\copy0\kern-\wd0
 \kern.05em\copy0\kern-\wd0
 \kern-.025em\raise.0433em\box0}
\def\boldvf{\pmb{$\!v(f)\,$}}
\def\boldf{\pmb{$\!f$\,}}

\begin{document}
\title{Phase-Space Structure \& Substructure of Dark Halos}
\author{Avishai Dekel}
\affil{Racah Inst. of Physics, The  Hebrew University, Jerusalem 91904, Israel}
\author{Itai Arad}
\affil{Institute of Astronomy, Madingley Road, Cambridge CB3 OHA, UK}

\begin{abstract}
  A method is presented for computing the 6-D phase-space density
  $f(\Bx,\Bv)$ and its PDF $v(f)$ in an N-body system.  It is based on
  Delaunay tessellation, yielding $v(f)$ with a fixed smoothing window
  over a wide $f$ range, independent of the sampling resolution. It is
  found that in a gravitationally relaxed halo built by hierarchical
  clustering, $v(f)$ is a robust power law, $v(f) \propto f^{-2.5 \pm
    0.05}$, over more than 4 decades in $f$, from its virial level to
  the current resolution limit.  This is valid for halos of different
  sizes in the $\Lambda$CDM cosmology, indicating insensitivity to the
  initial-fluctuation power spectrum as long as the small-scale
  fluctuations were not completely suppressed. By mapping $f$ in
  position space, we find that the high-$f$ contributions to $v(f)$
  come from the ``cold" subhalos within the parent halo rather than
  the halo central region and its global spherical profile.  The $f$
  in subhalos near the halo virial radius is more than 100 times
  higher than at the halo center, and it decreases gradually with
  decreasing radius.  This indicates phase mixing due to mergers and
  tidal effects involving puffing up and heating.  The phase-space
  structure provides a sensitive tool for studying the evolution of
  subhalos during the buildup of halos. One wishes to understand why
  the substructure adds up to the universal power law in $v(f)$.  It
  seems that the $f^{-2.5}$ behavior is related to the hierarchical
  clustering process and is not a general result of violent
  relaxation.
\end{abstract}

%%%%%%%%%%%%%%%%%%%%%%%%%%%%%%%%%%%%%
\section{Introduction}
\label{sec:intro}

% DM halos origin of structure not understood
Dark-matter halos are the basic entities in which luminous galaxies
form and live. They dominate the gravitational potential and have a
crucial role in determining the galaxy properties.  While many of the
systematic features of halo structure and kinematics have been
revealed by $N$-body simulations, the origin of these features is
still not understood, despite the fact that they are governed by
simple Newtonian gravity.

% example: density profile
The halo density profile $\rho(r)$ is a typical example.  It is found
in the simulations to have a robust non-power-law shape (originally
Navarro, Frenk \& White 1997, NFW; Power \etal 2003; Hayashi \etal
2004 and references therein), with a log slope of $-3$ at large radii,
varying gradually toward $-1$ or even flatter at small radii.  The
slope shows only a weak sensitivity to the cosmological model and the
initial fluctuation power spectrum (e.g. Colin \etal 2003; Navarro
\etal 2004), indicating that its origin is due to a robust relaxation
process rather than specific initial consitions.  In particular,
violent relaxation (Lynden-Bell 1967) may be involved in shaping up
the density profile, but we have no idea why this profile has the
specific NFW shape.

% example: kinematics
The properties of the velocity dispersion tensor is another puzzle.
The velocity dispersion profile is slightly rising at small radii and
slightly falling at large radii but is rather flat overall (Huss, Jain
\& Steinmetz 1999a; 1999b).  The profile of the anisotropy parameter
$\beta(r)$ indicates near isotropy at small radii that is developing
gradually into more radial orbits at large radii (Colin \etal 2000).
For a sperical system in equilibrium, the $\sigma(r)$ and $\beta(r)$
are related to $\rho(r)$ via the Jeans equation, but it is not at all
clear why $\sigma(r)$ or $\beta(r)$ have these specific shapes.

% Taylor & Navarro
An interesting attempt to address the origin of the halo profile has
been made by Taylor \& navarro (2001), who measured a poor-man
phase-space density profile by $f_{\rm TN}(r) =\rho(r)/\sigma(r)^3$,
and found that it displays an approximate power-law behavior, $f_{\rm
  TN} \propto r^{-1.87}$, over more than two decades in $r$.  Using
the Jeans equation, they showed that this power law permits a whole
family of density profiles, and that a limiting case of this family is
a profile similar to NFW, but with an asymptotic slope of $-0.75$ as
$r \rightarrow 0$.  This scale-free behavior of $f_{\rm TN}(r)$ is
intriguing, and it motivates further studies of halo structure by
means of phase-space density.

% hierarchical buildup
The simulations of the $\Lambda$CDM cosmology also reveal that the
halos are bulit by a rougly self-similar hierarchical clustering
process, where smaller building blocks accrete and merge into bigger
halos. At every snapshot, every halo contains a substructure of
subhalos on top of a smooth halo component that has been tidally
stripped from an earlier generation of substructure.  Some of the
important dynamical processes invloved in this hierarchical halo
buildup are understood qualitatively, including dynamical friction,
tidal stripping and mergeres.  However, a complete understanding of
how these processes work in concert to produce the halo structure and
kinematics is lacking.

% cusp by hierarchical buildup
Attempts have been made to explain an inner density cusp using  
toy models of dynamical stripping and tidal effects during the
halo buildup by mergers (e.g. Syer \& White 1998; Dekel, Devor \& Hetzroni
2003; Dekel \etal 2003). 
However, a similar halo density profile seems to be produced also
in simulations where substructure has been artificially suppressed
(Moore \etal 1999b; Alvarez, Shapiro \& Martel 2002), 
indicating that the process responsible for
the origin of this density profile might be a more robust feature of gravity.

% observations and other issues where substructure is relevant
The issue of halo substructure has become timely
both observationally and theoretically.
Tidal streams associated with dwarf satellite galaxies
are observed in the halos of the Milky Way and M31
and reveal their histories (Ibata \etal 2001; this proceedings).
Gravitational-lenses provide preliminary indicatons
for the presence of substructure in halos at the level predicted 
by the $\Lambda$CDM scenario (Dalal \& Kochaneck 2002).
In contrast, the observed number density of dwarf galaxies seems to be 
significantly lower, thus posing a ``missing dwarf problem" 
(Klypin \etal 1999; Moore \etal 1999a).
Also, the ``angular-momentum problem" of disk galaxies 
(e.g. Navarro \& Steinmetz 2000; Bullock \etal 2001a)
is probably associated with the evolution of substructure in halos
(Maller \& Dekel 2002; Maller, Dekel \& Somerville 2002).
While these problems necessarily invlove baryonic proceses, understanding 
the gravitational evolution of substructure is clearly a prerequisite
for solving them.

% f(x,v) 6D
Aiming at the origin of halo structure, we report here on a first
attempt by Arad, Dekel \& Klypin (2004, ADK) to address directly the
halo phase-space structure.  The fundamental quantity in the dynamical
evolution of gravitating systems is the full, 6D, phase-space density
$f(\Bx,\Bv)$, which intimately relates to the underlying Vlasov
equation, and lies behind any (violent) relaxation process that gives
rise to the virialized halo structure.  Ideally, one would have liked
to compute $f$ free of assumptions regarding spherical symmetry,
isotropy, or any kind of equilibrium, but computing densities in a 6D
space is a non-trivial challenge.  The state-of-the-art N-body
simulations, with more than million particles per halo, allow for the
first time an attempt of this sort. ADK developed a succesful
algorithm for measuring $f(\Bx,\Bv)$, and studied its relevant
properties and the associated systematic and random uncertainties.
They then applyied this algorithm to simulated virialized halos in the
$\Lambda$CDM cosmology, and obtained two surprising new results.
First, the phase-space volume 
distribution of $f$ is a universal power-law, valid in all virialized
halos that form by hierarchical clustering. Second, this power law is
not directly related to the overall density profile, but is rather
driven by the halo substructure.  We thus learn that $f(\Bx,\Bv)$
provides a useful tool for studying the hierarchical buildup of
dark-matter halos and the evolution of substructure in them.

%%%%%%%%%%%%%%%%%%%%%%%%%%%%%%%%%%%%%%%%
\section{More about \boldf\ and \boldvf}
\label{sec:vf}

% def
A collisionless system is completely determined by the fine-grained
phase-space density function $f(\Bx,\Bv,t)$.  The evolution of $f$ is
governed by the Vlasov equation,
\begin{equation}
  \label{eq:vlasov}
  \partial_t f + \Bv \cdot \nabla_{\Bx} f 
    - \nabla_{\Bx}\Phi \cdot \nabla_{\Bv} f = 0 \ , 
\end{equation}
with $\Phi(\Bx)$ the gravitational potential, related
self-consistently to $f(\Bx,\Bv)$ by the Poisson's integral
\begin{equation}
  \Phi(\Bx) = -G\int \!\! d\Bx' d\Bv \, 
                \frac{f(\Bx',\Bv)}{|\Bx-\Bx'|} \ .
\end{equation}
We therefore assume that a true understanding of the nature of
self-gravitating collisionless systems must involve $f(\Bx,\Bv)$ as a
primary ingredient. 

While the 6D $f(\Bx,\Bv)$ is hard to deal with,
there is a simpler function, $v(f)$, which is intimately related 
to $f(\Bx,\Bv)$, yet is much simpler to handle. It is defined as
\begin{equation}
\label{def:vf}
v(f=f_0) \equiv \int \!\! d\Bx d\Bv\, \delta\big[ f(\Bx, \Bv, t) - f_0\big] \ ,
\end{equation}
such that $v(f)df$ is the volume of phase-space occupied by
phase-space elements whose density lies in the range $(f,f+df)$. 
The Valsov equation ensures that every phase-space elements preserves 
its density along its path, as therefore $v(f)$ is conserved.        

% coarse-grained and mixing
However, as the system evolves, phase-space patches of high $f$ are 
stretched and spiral into regions with low $f$, and 
become thiner such that $f$ is varying over increasingly smaller
scales.  At some point one can no longer measure $f$ but rather an
average of it over some finite volume, referred to as the
``coarse-grained" phase-space density, which we denote $\cf$.  The
coarse-grained $\cv(\cf)$ is no longer conserved.  In the course of a
collapse or merger of a dark-matter halo, rapid global fluctuations of
the gravitational potential re-distribute the energies of each
phase-space element and lead to mixing.  After a few global dynamical
times, the potential fluctuations fade away, $\cf$ stabilizes, and it
can be viewed as the ``physical" phase-space density of the system,
since the microscopic fluctuations of $f$ can no longer be measured or
affect the gravitational potential. 

One obvious constraint on the final $\cf$, as an average of $f$, is
that their maximum values must obey $\bar{f}_{\rm max} \leq f_{\rm
  max}$.  Since the initial $\cv(\cf)$ is identical to $v(f)$, if the
initial $\cv(\cf)$ vanishes for $\cf>f_{\rm max}$, then so does the
final $\cv(\cf)$.  There are several additional constraints imposed by
the given $v(f)$ on the final $\cv(\cf)$, specified by the
\emph{mixing theorem} (Tremaine, H\'{e}non \& Lynden-Bell 1986), whose
strength is rather limited because it only provides an
integro-differential inequality constraint on $\cv(\cf)$.

%\subsection{$v(f)$ versus $\rho(r)$}
In general, the same $v(f)$ can describe 
for different systems. However, if the system is spherically symmetric
and stationary, such that $f$ is a function of the energy alone,
$f(\Bx,\Bv) = f(\epsilon)$, then there is a unique relation between
$v(f)$, $f(\epsilon)$ and $\rho(r)$.  In particular, using dimensional
arguments and the virial theorem, one can show that if $\rho(r)
\propto r^{-\alpha}$, then $v(f) \propto f^{-\beta}$ with
\begin{equation}
\label{eq:hi-connection}
\alpha = \frac{18-6\beta}{4-\beta} \ .
\end{equation}
Therefore, the values in the range $0 \leq \alpha \leq 2$, which are
relevant for the inner regions of halos where $f$ is high, correspond
to a narrow range of $\beta$ values, $3 \geq \beta \geq 2.5$.  The
value $\beta=2.5$ corresponds to the singular isothermal sphere
$\alpha=2$, while $\beta=2.8$ corresponds to an $\alpha=1$ cusp, and
$\beta=3$ corresponds to a flat core, $\alpha=0$.

%%%%%%%%%%%%%%%%%%%%%%%%%%%%%%%
\section{Measuring \boldvf\ in an $N$-body System}
\label{sec:computing}

% Difficulties in a straightforward box-counting.
We wish to measure the 6D $f(\Bx,\Bv)$ of a system represented by $N$
particles of mass $m$ each.  Counting particles in uniform cells is
impractical because even with $10^6$ cells there would be only 10
cells along each axis.  We therefore use an adaptive grid, where the
cells vary in size and in shape to allow a proper resolution where
needed.  A particularly robust method of this type is based on the
\emph{Delaunay Tessellation} (Delaunay 1934), which has already been
implemented in a cosmological context in 3D (Bernardeau \& van de
Weygaert 1996; Schaap \& van de Weygaert 2000).

% Delaunay
A tessellation is the division of $R^d$ space into a complete covering
of mutually disjoint convex polygons.  
In the Delaunay tessellation, 
every $d+1$ points whose circumsphere [i.e., the $(d-1)$-dimensional
sphere that passes through all of them] does not encompass any other
point define a $d$-dimensional polyhedron which makes a \emph{Delaunay
  Cell}.  ADK followed van de Weygaert (1994) in using the algorithm
by Tanemura, Ogawa \& Ogita (1983). An analysis of a halo with $10^6$
particles, which lasts about a week on a standard CPU, produces $\sim
10^9$ cells. A typical particle is surrounded by $\sim 7,000$ cells,
made out of $\sim 200$ neighboring particles.

% Recovering $f$ and $v(f)$ from the Tessellation}
The value of $f$ at particle $i$ is estimated by
$f_i \equiv (d+1){m}/{V_i}$,
where $V_i$ is the total volume of all the cells neighboring to particle $i$,
$d$ is the dimensionality (here $d=6$),
and the factor $d+1$ ensures mass conservation.
In order to estimate $f$ at any point $(\Bx,\Bv)$, 
one averages over the $f_i$'s of the seven particles that define the cell.
The desired $v(f)$ is obtained by first computing
its cumulative version $V(f)$,
\begin{equation}
V(f_0) = \int_{f_0}^\infty \!\! v(f') \, df' 
= \int_{f(\Bx, \Bv) > f_0}\!\!\!\!\! d\Bx d\Bv 
\to \sum_{f_\nu > f_0} |D_\nu| \ , 
\end{equation}
and then differentiating $v(f) = -{dV(f)}/{df}$.  
The sum in the above formula is of all volumes of Delaunay cells 
$D_\nu$, whose average $f_\nu$ is greater than $f_0$.            
In what follows, we sometimes denote the Delaunay-measured $f$ and
$v(f)$ by $\fdel$ and $\vdel$.

%\subsection{Error Estimate}
When measuring the $v(f)$ of a cosmological $N$-body system, one
expects two types of errors.  First, the errors in the underlying $f$
associated with errors in the numerical simulation itself, e.g., due
to two-body relaxation effects, force estimation, or time integration.
A way to estimate these errors is by re-simulating the same system
with different codes and with different sets of numerical parameters.
A systematic testing of this sort will be reported elsewhere (Arad,
Dekel \& Stoehr, in prep.).  Meanwhile, ADK compared several different
halos simulated with different resolutions and with different codes,
and found that all the halos tested recover almost the same $v(f)$.

The other type of errors originate from the fact that we estimate a
smooth $f$ from a finite set of particles using a specific adaptive
technique. Here one may encounter both statistical and systematic
errors. Some of these errors would decrease as the number of particles
is increased, whereas other errors are an inherent part of the method.
ADK introduced a simple statistical model based on Varonoi
tesselation, which resembles the Delaunay technique and yet lends
itself more easily to analytical treatment. The predictions of this
model were then tested against numerical experiments with synthetic
systems.

It is found that $\fdel$ at each particle is drawn from a 
probability distribution function of $\fdel/\ftrue$. 
In a typical realization with $10^6$ 
particles, the width of this distribution is about one decade,
describing the typical fluctuations of $\fdel$ about $\ftrue$. 
As the number of particles $N\to\infty$, the shape
of the distribution approaches an asymptotic limit, with a
\emph{finite} width of about one quarter of a decade, 
implying local fluctuations even in the infinite limit.

It is then realized that $\vdel(f)$ can be viewed as a convolution of 
the true $v(f)$ and a fixed window function,
\begin{equation}
\vdel(f=f_0)=\int_0^\infty\!\! \vtrue(f) f^{-1}_0 p(f_0/f)\,df \ .
\end{equation}
If $v(f)$ is close to a power-law, the
difference between $\vdel(f)$ and $\vtrue(f)$ 
is negligible over a large range of scales.
This is a very useful feature of the method.
      
The relative statistical error in $\vdel(f)$ is proportional
to $1/\sqrt{N}$, and can be approximated by
\begin{equation}
\Delta f \simeq  \left( {m \over f \left<\Vdel(f)\right>} \right)^{1/2} \ ,
\label{eq:cal-stat-error}
\end{equation}
with $\Vdel(f) = \int_f^\infty \!\!\vdel(f')\,df'$ and $m=M/N$. 
In practice, this means that when $N \geq 10^6$,
the statistical error is
negligible for a very wide range of $f$. Moreover, in regions
where there are large statistical errors, these errors are usually
overwhelmed by systematic errors.

%%%%%%%%%%%%%%%%%%%%%%%%%%%%%%%%%%%%%%%%%%%%%%%
\section{A Universal Scale-Free \boldvf}
\label{sec:universal}

The $v(f)$ is measured from several different halos, in three
different mass ranges, simulated within the $\Lambda$CDM cosmology
with two different $N$-body codes.  We focus on virialized halos out
to slightly outside the virial radius.  The value of $f$ that can be
crudely associated with the virial radius of a given halo, where one
may expect a qualitative change in the behaviour of $v(f)$, is
estimated by $\fvir = \pi^{-3/2} \rho/\sigma^3$, with $\rho$ and
$\sigma$ the mean virial quantities.  On the other side, the
reliable-measurement range is marked by $f=\f02$, below which the
statistical error in $v(f)$ due to the DTFE is below $20\%$ according to 
%\equ{cal-stat-error},
eq.~7.  This formula was verified by the $v(f)$ of mock systems 
with $10^5$ particles.                                 
The statistical error is expected to be practically negligible in the
range $\fvir < f < \f02$.  When tested using synthetic datasets for
which the true $v(f)$ is known, we learn that with $10^5$ particles
the method recovers the true $v(f)$ very well over a range of 5-6
decades in $f$, while with $10^6$ particles the range spans $\sim 10$
decades.

%\subsection{ART Simulations on Different Scales}
The results described in ADK are based on three different cosmological
simulations, two using the ART code (Kravtsov, Klypin \& Khokhlov
1997), and the third using the TPM code (Bode, Ostriker \& Xu 2000;
Bode \& Ostriker 2003).  The assumed cosmological model is the
standard $\Lambda$CDM with $\Omega_{\rm m} = 0.3$, $\Omega_\Lambda =
0.7$ and $h=0.7$ today.  Halos were sampled from the simulatied
periodic boxes of sides $L=1$, $25$ and $320\hmpc$, which we denote
$L1$, $L25$ and $L320$. 
The simulations are by Colin \etal (2003), by Klypin \etal (2001), and
by Wambsganss, Bode \& Ostriker (2004) and Weller, Bode \& Ostriker
(2004).  The force resolution is $87$, $140$ and $4700 {\rm pc}$.  The
particle mass is $7\times 10^3$, $1.2 \times 10^6$ and $2.6 \times
10^9 \msun$.  The halo masses analyzed from these simulations
correspond to dwarf galaxies ($10^9-10^{10}\msun$), normal galaxies
($\sim 10^{12}\msun$) and clusters of galaxies ($\sim 10^{15}\msun$)
respectively.  The ART halos were sampled by $\sim 10^6$ particles
within $\sim 1.1 \Rvir$, while the TPM halos were sampled by only
$4\times 10^5$ particles within the corresponding radius.  The L1
halos were analyzed at $z=2.33$.  More details about the simulations
and the halo properties are summarized in Table 1 of ADK.  The $v(f)$
curves for these nine halos are shown in 
%\Fig{vf}.
Fig.~1

\begin{figure}
\vskip 7.0cm
{\includegraphics{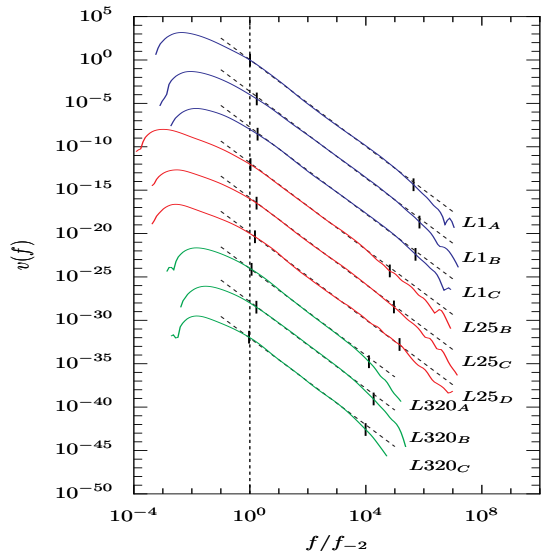}}
\vskip 0.2cm
\caption{The volume distribution of phase-space density, $v(f)$,
for each of the nine halos analysed in ADK.
The curves were shifted to coincde at $f=f_{-2}$,
where the local log slope of $v(f)$ is $-2$,
and were then shifted vertically by 4 decades relative to each other.
A power-law line $v(f)\propto f^{-2.5}$ is shown on top of each curve.
Marked on each curve are the virial-radius level $\fvir$ and the 20\%
statistical error limit $\f02$.}
\label{fig:vf}
\end{figure}

The figure shows that in every halo $v(f)$ is well fit by a power-law,
\begin{equation}
v(f) \propto f^{-2.50\pm0.05} \ ,
\end{equation}
over 3 to 5 decades in $f$. It is typically valid between about
$\fvir$ and slightly below $\f02$.  
Outside this range, $v(f)$ gradually 
deviates downward.  In the low-$f$
regime the deviation is associated with departure from the virial
regime, while the high-$f$ deviation is consistent with being due to the
limited mass resolution, 
as indicated by $\f02$ and by the error analysis of ADK.
The high-$f$ deviation from the power-law tends to occur at a smaller $f$
value in L25, and even smaller in L320, due to the fact that $\Nvir$ is
smaller respectively.

The power law in $v(f)$ does not show a significant dependence on 
halo mass. There may be a marginal trend for slight steepening of
$v(f)$ as a function of mass, but only from steeper than $f^{-2.45}$
at $\sim 10^{9}\msun$ to flatter than $f^{-2.55}$ at $\sim 10^{15}\msun$.
This indicates relative insensitivity to the exact slope of the initial
fluctuation power spectrum, which varies across the range from dwarf galaxies
to clusters of galaxies. Additionally, the fact that we obtained
essentially the same $v(f)$ from simulations using two different
numerical codes, indicates that the shape of $v(f)$ is not an
artifact of a particular simulation technique.

%%%%%%%%%%%%%%%%%%%%%%%%%%%%%%%%%%%%%
\section{Substructure}
\label{sec:substructures}

%\subsection{Clumpiness in Phase-Space Density}

% spherical case: rho(r) vs v(f)
When $f(\Bx,\Bv)$ is function of the energy alone, and the halo is
spherical and isotropic, the power-law $v(f) \propto f^{-2.5}$ implies
via
%\equ{hi-connection} 
eq.~4 that the real-space density profile must also be a power law, in
fact an isothermal sphere $\rho(r) \propto r^{-2}$, at least over some
finite range in $r$.  This is clearly not the case
%(\se{intro}), 
(\S1), 
indicating that $f$ is not a function of energy alone, and the system
must deviate from spherical symmetry or isotropy.  This could be due
to the clumpy substructure of the halo, where the surviving subhalos
contribute high phase-space densities to $v(f)$, thus making it
shallower than expected from a smooth system with an inner density
slope flatter than $-2$.
             
% maps
%\Figu{denmapB} 
Figure 2 shows density and phase-space density maps of an equatorial
slice from one of the halos, in which the color represents the
densities $\rho$ and $f$ averaged over the values assigned to each
particle in a small real-space volume (of side $\sim 0.005 \Rvir$).
The $\rho$ of each particle was calculated using a 3D Voronoi
tessellation (van de Weygaert 1994), which is similar in its adaptive
nature to the Delaunay tessolation used to estimate $f$.  While the
real-space density maps are dominated by the familiar smooth trend of
density decreasing outward, perturbed by several tight clumps
throughout the halo, the global trend with radius becomes much less
prominent in the phase-space density maps, with the subhaloes
contributing the highest peaks, especially in the outer regions of the
halo.  Moderate $f$ peaks are found everywhere (reddish yellow), and
the very high peaks (bright yellow) are preferentially found in the
periphery.  The central peak in $f$ is quite modest in comparison; the
elongated structures near the center of the shalo are most likely
merging subhalos.

\begin{figure}
\vskip 6.0cm
{\includegraphics{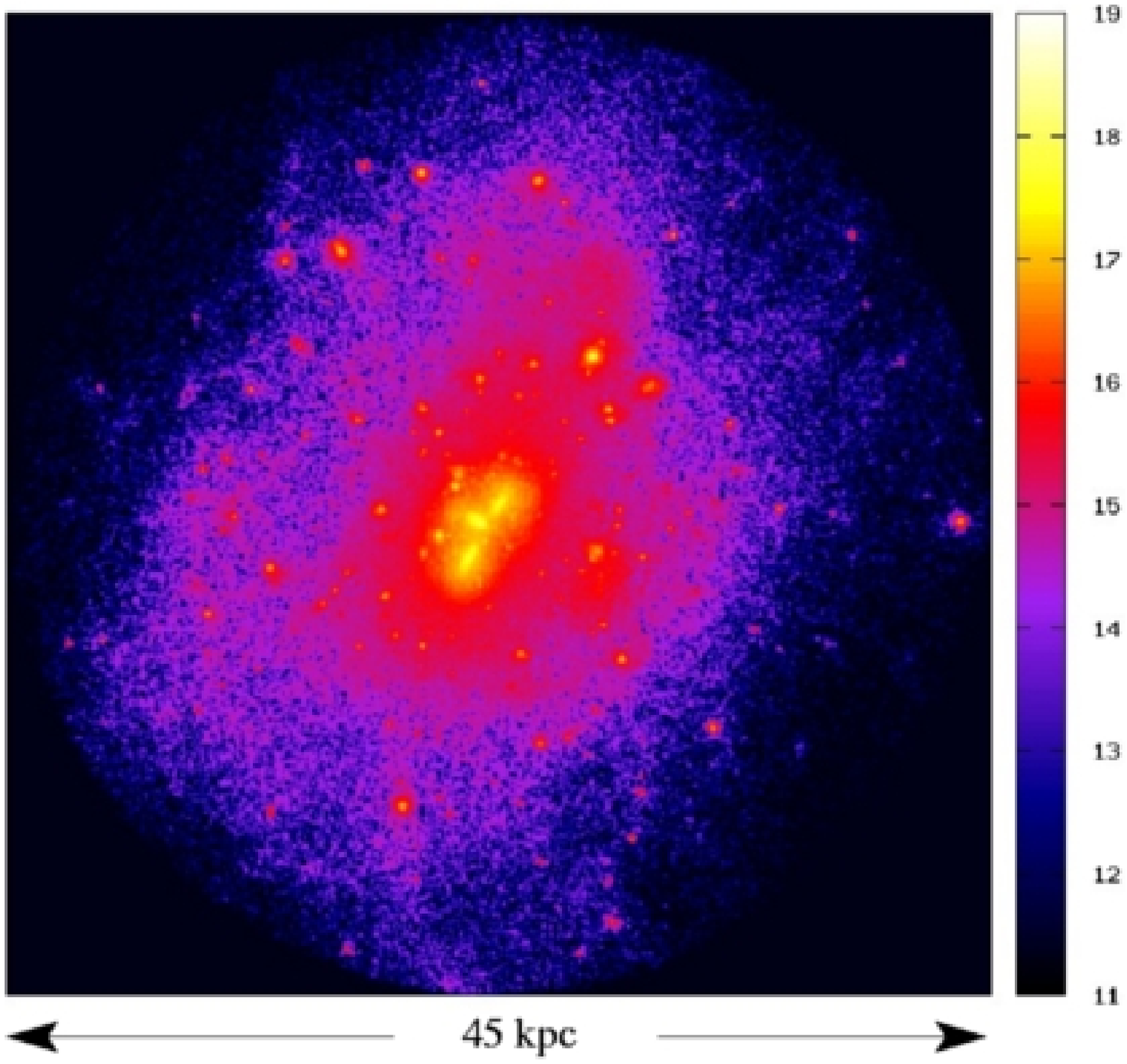}}
{\includegraphics{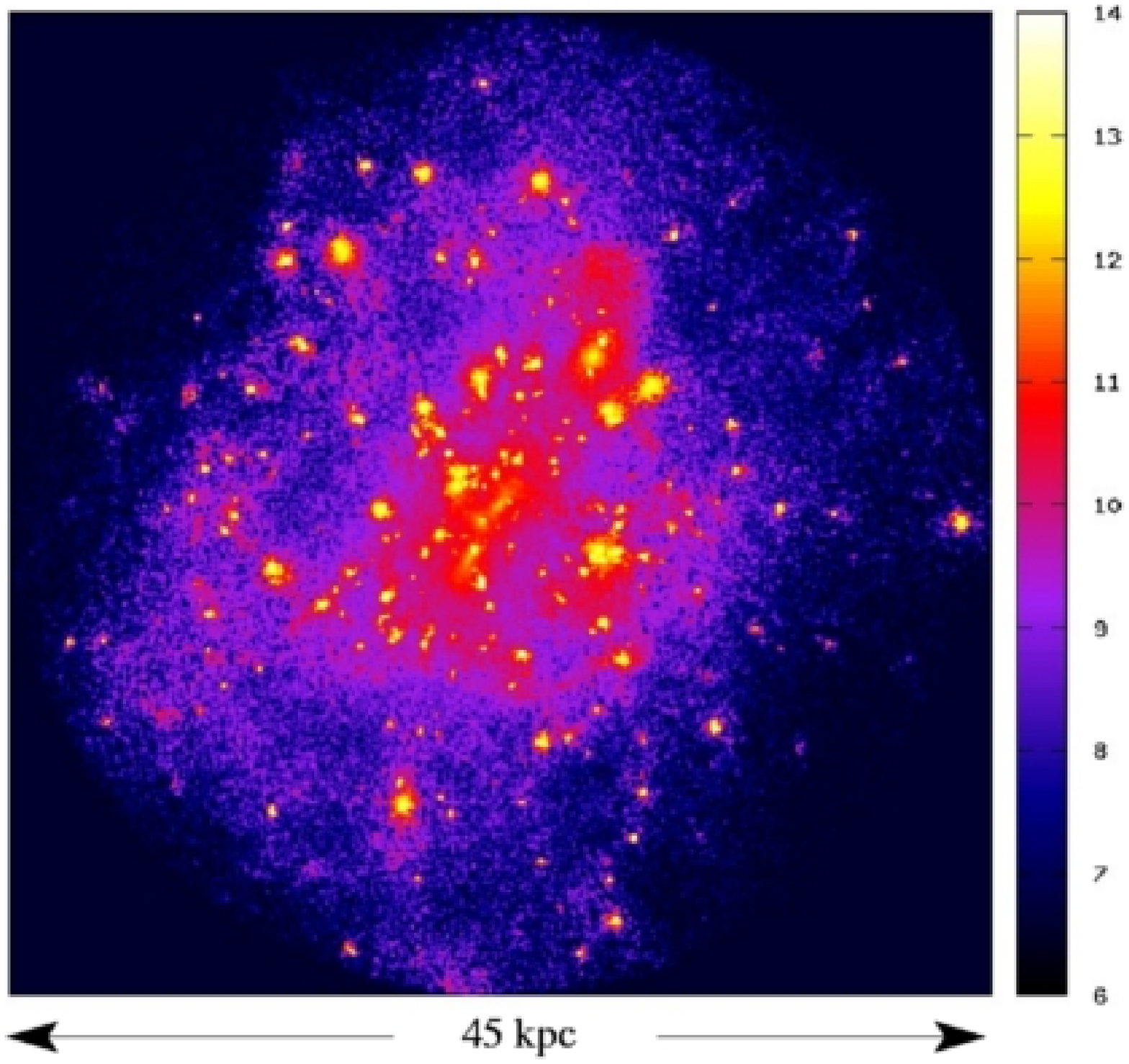}}
\vskip 0.2cm
\caption{Density maps of dwarf halo $L1_B$ in a slice
  of thickness $0.4 R_{\rm vir}$.  Left: real-space density.  Right:
  phase-space density.  The units in the colour key are
  $\log(\rho/[\msun Mpc^{-3}])$ and $\log(f/[\msun
  Mpc^{-3}\mbox{km}^{-3}s^3])$ respectively. The very-high $f$ values
  are found inside clumps which are typically far away from the halo
  center.}
\label{fig:denmapB}
\end{figure}

% f(r) spikes
%\Figu{fr} 
Figure 3 shows $\rho$ and $f$ associated with a random subset of the
N-body particles as a function of their distance $r$ from the halo
center.  A large portion of these particles follow the global trend of
decreasing density with radius -- they could be associated with a
smooth-background component, for which $f$ is approximately a function
of energy alone.  At radii $r>1\kpc$, the high-$f$ values come in as
``spikes" corresponding to the subhalos.  While the spikes in $\rho$
reach values comparable to the central peak, the spikes in $f$ could
be more than 100 times higher, indicating that the subhalos are both
\emph{compact} and \emph{cold}.

% tidal heating
The spikes seem
to be lower and broader as they get closer to the halo center, and
they completely blend into the smooth background inside $\sim 2\kpc$.
This indicates that the subhalos phase-mix and lose their high
phase-space densities as they approach the halo center.  This seems to
be the natural result of mergers and \emph{tidal effects}, which 
puff up the subhaloes and especially heat them up.

\begin{figure}
\vskip 5.0cm
{\includegraphics{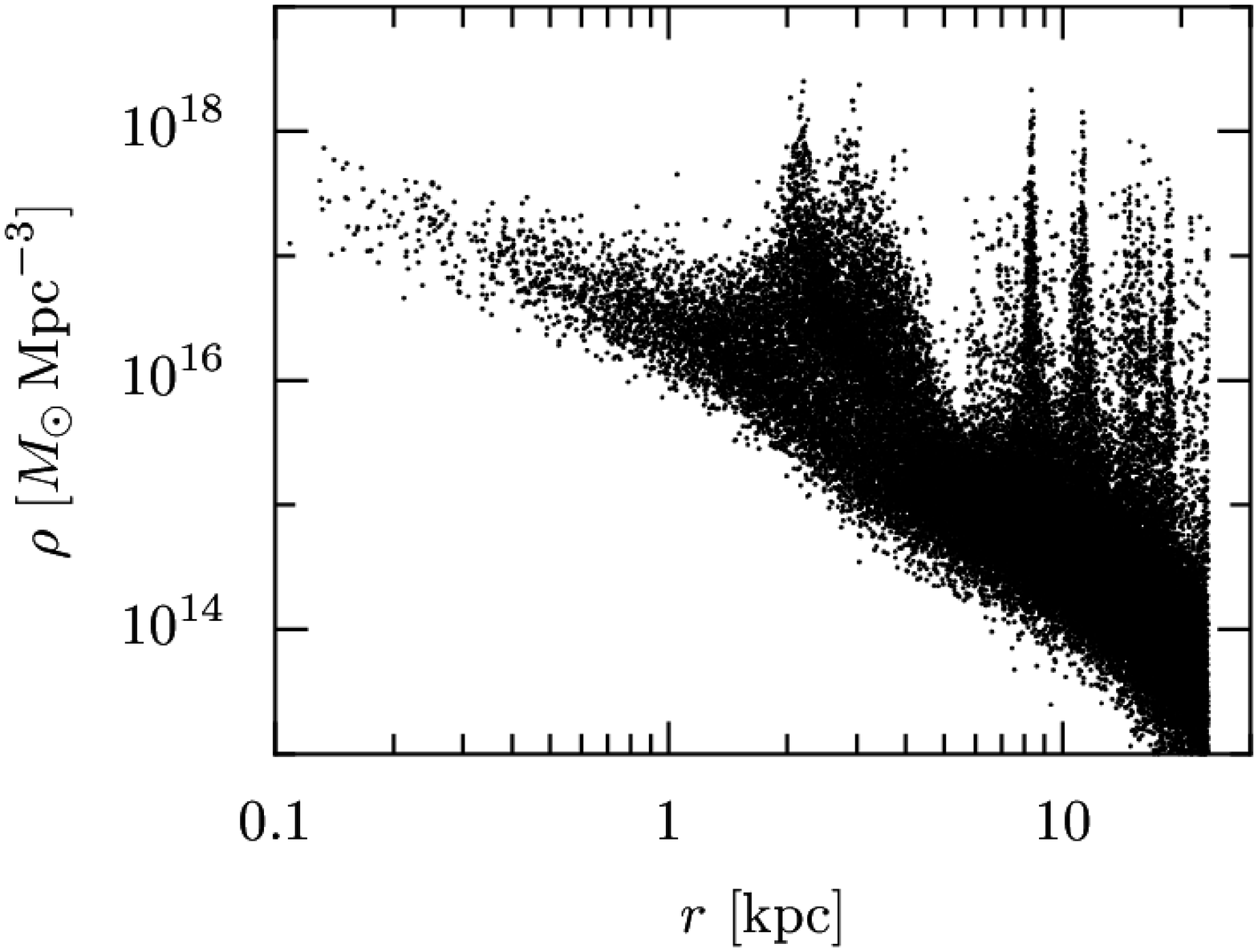}}
{\includegraphics{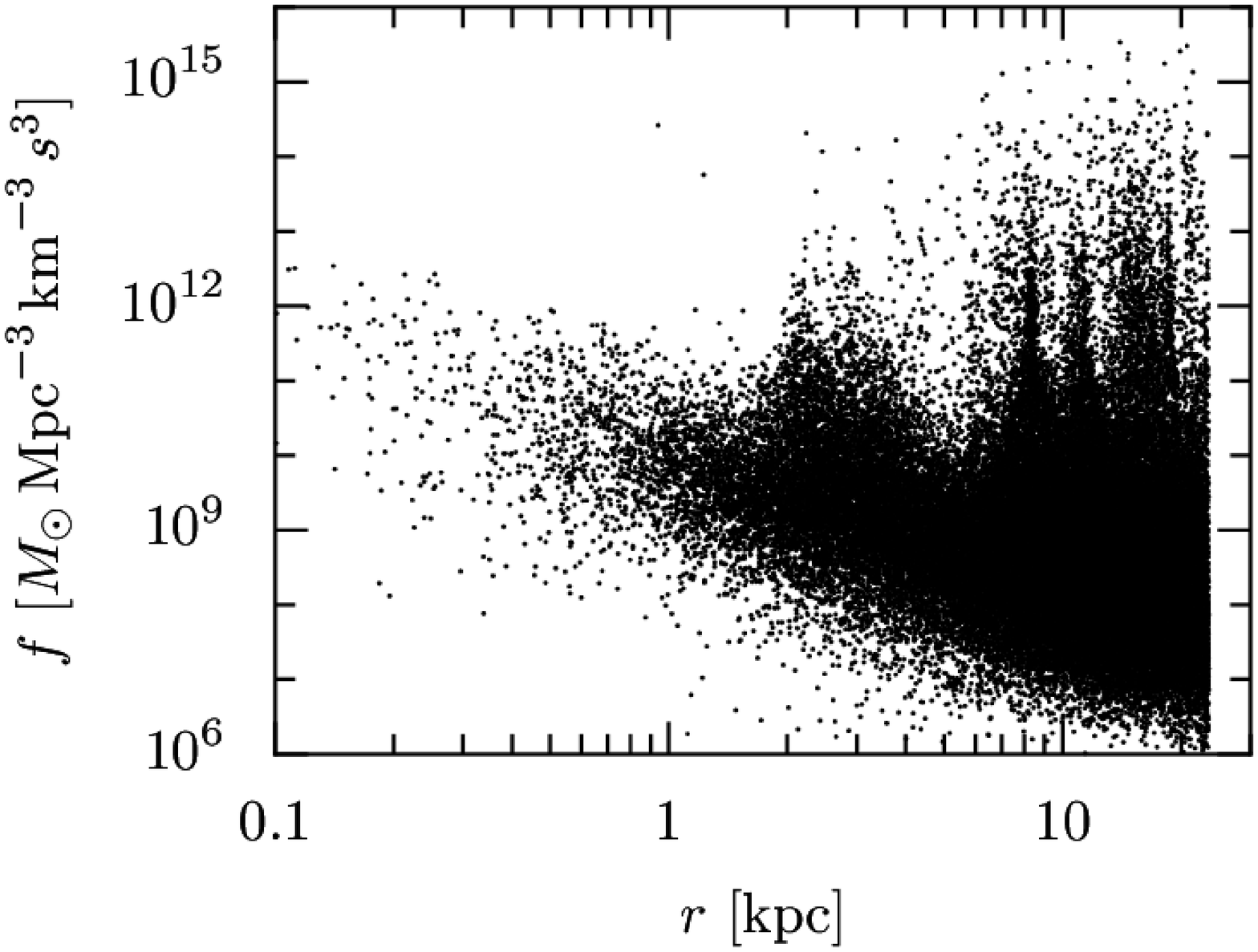}}
\vskip 0.2cm
\caption{Densities as a function of radius
for dwarf halo $L1_B$, using a random set of 4\% of the particles.
Left: real-space density.  Right: phase-space density.
The background particles define a general trend of decreasing density
with radius, while the spikes correspond to subhalos.
The phase-space density spikes are higher than the central peak
because the subhalos are cold. They become shorter and broader
at smaller radii, indicating heating by tidal effects and mergers.}
\label{fig:fr}
\end{figure}

%\subsection{Toy Model: Adding Up Small Haloes}
As a first attempt at trying to understand how the power law $v(f)
\propto f^{-2.5}$ in a halo of mass $M$ may originate from its
substructure, one may simply add up the typical contributions from the
general population of halos of different masses $m$ smaller than $M$,
as predicted in the $\Lambda$CDM cosmology.  Based on cosmological
N-body simulations (e.g., Moore \etal 1999a; Ghinga \etal 2000; De
Lucia \etal 2004), and in accordance with the Press-Schechter (1974)
approximation, the mass function of small-mass halos can be
approximated by $dn/dm \propto m^{-\gamma}$, with $\gamma \simeq
1.8-2.0$.  Supported by the simulations, one assumes that the average
density profiles of halos of different masses are simply scaled
versions of each other (e.g., Navarro \etal 1996) If all the halos
form at the same time, they have the same characteristic real-space
density $\rho_m=\rho$, so their typical radii scale like $r_m \propto
m^{1/3}$.  Based on the virial theorem, the velocity dispersions then
scale like $\sigma_m \propto m^{1/3}$.  Therefore, the typical
phase-space volume of a halo of mass $m$ scales like $V_m \propto
r_m^3 \sigma_m^3 \propto m^2$, and its typical phase-space density is
$f_m \propto m/V_m \propto m^{-1}$.  With this scaling one can show
that $v(f)$ should be flatter than $v(f) \propto f^{-(4-\gamma)}$.
For $\gamma \geq 1.8$, this means that $v(f)$ is shallower than
$f^{-2.2}$, which is significantly shallower than the measured
$f^{-2.5\pm0.05}$.  Including the fact that small halos form first in
a $\Lambda$CDM scenario makes a negligible difference.  One can
conclude that the population of subhalos within a bigger host halo
must be different than the general population of halos.  As a result
of mergers and tidal effects, the subhalos are expected to have a
different mass function, their shape properties are likely to vary
differently with mass, and both effects probably vary with radius.
Thus, the phase-space density is likely to provide a useful tool for
studying the dynamical evolution of subhalos in their parent halos.

%%%%%%%%%%%%%%%%%%%%%%%%%%%%%%%%%%%%%
\section{Discussion}
\label{sec:conc}

It is important to verify that these results are not numerical
artifacts. 
Based on the error analysis and tests with mock datasets, we believe 
that the $v(f)$ measured by the DTFE algorithm genuinely reflects the
true phase-space properties of the given $N$-body system over a broad
range of $f$. The question is whether the phase mixing suffered by the
subclumps is an artifact of numerical effects such as few-body
relaxation, leading to underestimated inner densities and/or
overestimated internal velocities (Binney 2003).  The apparent
agreement between simulations run with different codes and different
resolutions is encouraging. In order to specifically address the
effect of two-body 
relaxation, we intend to run twice a simulation of the same halo with
the same number of particles but with a different force resolution
(ongoing work with F. Stoehr).

Assuming that the simulations genuinely reflect the true physical
behaviour, the origin of the robust power-law shape of $v(f)$ from the
merging substructure becomes a very interesting theoretical issue.  As
demonstrated in
%\se{substructures}, 
\S5, a simple model using the mass function and the scaled profiles of
the general halo population in the $\Lambda$CDM scenario does not
reproduce the correct power law.  This, and the apparent trend of the
$f$ spikes with radius, indicate that the structural and kinematical
evolution of the subhaloes in the parent halo are important. Studies
of tidal heating and stripping may be found useful in this modelling.

It would be interesting to follow the phase-space evolution and the
contribution to the overall $v(f)$ by a single, highly resolved
subhalo, or many of those, as they orbit within the parent halo and
approach its center. This may help us understand the nature of the
interaction between the parent halo and its subhaloes, and the origin
of the $v(f)$ power law (ongoing works with E. Hayashi and with B.
Moore).

% time evolution - replacing a subsection
We saw that the power-law behavior of $v(f)$ is limited to the virial
regime.  It would be interesting to learn how this shape evolves in
time as the halo virializes.  A preliminary study (to be concluded and
reported in another paper) indicates that in the intermediate-$f$
regime the $v(f)$ of a pre-virialized system is significantly flatter
than $f^{-2.5}$, while in the high-$f$ regime it drops in a much
steeper way.  The $f^{-2.5}$ behavior seems to be a feature unique to
virialized systems.

% hierarchical vs monolitic - replacing a subsection
We learned that in the haloes that are built by hierarchical clustering,
the power-law behavior $v(f)\propto f^{-2.5}$ reflects the
halo substructure. It would be interesting to find out whether
this power-law behavior actually requires substructure, or it is
a more general phenomenon of virialized gravitating systems,
valid independently of substructure.  One way to answer this question
would be to analyse simulated haloes in which all fluctuations of
wavelengths smaller than the halo scale were removed, resulting in
a smooth halo formed by monolithic collapse, 
with no apparent substructure in the final configuration.
As described in 
%\se{intro},
\S1,
such haloes are known to still have NFW-like density profiles in real space,
and one wonders whether they also have the magic power-law $v(f)$.
There are preliminary indications for a steeper $v(f)$ in this case
(Arad, Dekel \& Moore, in preparation). 
If confirmed, it would indicate that the $f^{-2.5}$ behavior, while
insensitive to the exact slope of the initial power spectrum, 
is unique to the hierarchical clustering process, and is not a general
result of violent relaxation. 

Our current results are just first hints from what seems to be a
promising rich new tool for analysing the dynamics and structure of
virialized gravitating systems. The analysis could become even more
interesting when applied to haloes including the associated gaseous
and stellar components.

\acknowledgments{We thank Paul Bode, Stefan Gottloeber, Eric Hayashi,
Anatoly Klypin, Ben Moore, Julio Navarro and Felix Stoehr 
for their ongoing collaborations on the various parts of this projet.
We acknowledge stimulating discussions with Stefan Colombi, Donald
Lynden-Bell, Gary Mamon, Jerry Ostriker and Simon White.
IA is a Marie Currie Fellow.
AD is a Miller Visiting Professor at UC Berkeley.
This research has
been supported by the Israel Science Foundation grant 213/02, by
the German-Israel Science Foundation grant I-629-62.14/1999, and by NASA
ATP grant NAG5-8218.}

%%%%%%%%%%%%%%%%%%%%%%%%%%%%%%%%%%%%%%%
%\begin{thebibliographa}{11}


\begin{references}

\def\apj{ApJ}
\def\apjl{ApJL}
\def\apjs{ApJS}
\def\mnras{MNRAS}
\def\aap{A\&A}

\def\bibitem{\reference}

\bibitem
%[Alvarez, Shapiro \& Martel (2002)]{ref:Alvarez02}
Alvarez M.A., Shapiro P.R., Martel H., 2002, AAS, 200, 4103

\bibitem %[Arad, Dekel \& Klypin (2004)]{ref:Arad04}
Arad I., Dekel A., Klypin A.A., 2004, MNRAS, submitted, astro-ph/0403106
(ADK).

\bibitem %[Bernardeau \& van de Weygaert (1996)]{ref:Ber96}
Bernardeau F., van de Weygaert R., 1996, \mnras, 279, 693

\bibitem %[Binney (2003)]{ref:Bin03} 
Binney J., 2003, MNRAS, astro-ph/0311155

\bibitem %[Bode et al (2000)]{ref:Bode00} 
Bode P., Ostriker J.P., Xu G., 2000, \apjs, 128, 561

\bibitem %[Bode \& Ostriker (2003)]{ref:Bode03} 
Bode P., Ostriker J.P., 2003, \apjs, 141, 1

\bibitem %[Bullock et al. (2001a)]{ref:Bul01a} 
Bullock J.S., Dekel A., Kolatt T.S., Kravtsov A.V., Klypin A.A., Porciani C., 
Primack J.R., 2001a, ApJ, 555, 240

\bibitem %[Bullock et al. (2001b)]{ref:Bul01b} 
Bullock J.S., Kolatt T.S., Sigad Y., Somerville R.S., Kravtsov A.V., 
Klypin A.A., Primack J.R., Dekel A., 2001b, MNRAS, 321, 559

\bibitem %[Col\'{i}n et al (2000)]{ref:Col00} 
Col\'{i}n P., Klypin A.A., Kravtsov A.V., 2000, \apj, 539, 561

\bibitem %[Col\'{i}n et al (2003)]{ref:Col03} 
Col\'{i}n P., Klypin A.A., Valenzuela O., Gottl\"{o}ber S., 2003, ApJ,
submitted, astro-ph/0308348

\bibitem %[Dalal \& Kochanek (2002)]{ref:Dal02} 
Dalal N., Kochanek C.S., 2002, \apj, 572, 25

\bibitem %[Dekel, Devor \& Hetzroni (2003)]{ref:ddh03} 
Dekel A., Devor J., Hetzroni, G., 2003, MNRAS, 341, 326.

\bibitem %[Dekel et al. (2003)]{ref:Dek03} 
Dekel, A., Arad, I., Devor, J., Birnboim, Y., 2003, \apj, 588, 680
\bibitem
%[Delaunay (1934)]{ref:Del34} 
Delaunay B. V., 1934, Bull. Acad. Sci (VII) Classe Sci. Mat., 793

\bibitem %[De Lucia et al. (2004)]{ref:Luc04} 
De Lucia G.,  Kauffmann G., Springel V.,  White S.D.M.,  Lanzoni B.,  
Stoehr F.,  Tormen G.,  Yoshida N, 2004, \mnras, 348, 333

\bibitem %[Ghigna el al (2000)]{ref:Ghi00} 
Ghigna S., Moore B., Governato F., Lake G., Quinn T., Stadel J., 2000,
\apj, 544, 616

\bibitem %[Hayashi et al (2004)]{ref:Hay04}
Hayashi E., Navarro J.F., Power C., Jenkins A., Frenk C.S., White D.M., 
Springel V., Stadel J., Quinn T.R., \mnras, submitted, astro-ph/0310576

\bibitem %[Huss et al (1999a)]{ref:Hus99a} 
Huss A., Jain B., Steinmetz M., 1999a, \apj, 517, 64 % universality+dispersion

\bibitem %[Huss et al (1999b)]{ref:Hus99b} 
Huss A., Jain B., Steinmetz M., 1999b, \mnras, 308, 1011 %universal+dispersion

\bibitem %[Ibata et al (2001a)]{ref:Iba01} 
Ibata R., Irwin M., Lewis G.F., Stolte A., 2001, \apj, 547, 133

\bibitem%[Lynden-Bell (1967)]{ref:Lyn67} 
Lynden-Bell D., 1967, \mnras, 136, 101

\bibitem%[Klypin et al. (1999)]{ref:Kly99b} 
Klypin A.A., Kravtsov A.V., Valenzuela O., Prada, F., 1999, ApJ, 522, 82

\bibitem%[Klypin et al (2001)]{ref:Kly01} 
Klypin A.A., Kravtsov A.V., Bullock J.S., Primack J.R., 2001, \apj, 544, 903

\bibitem%[Kravtsov, Klypin \& Khokhlov (1997)]{ref:Kra97} 
Kravtsov A.V., Klypin A.A., Khokhlov A.M., 1997, \apjs, 111, 73

\bibitem%[Maller \& Dekel (2002)]{ref:Mal02} 
Maller A.H., Dekel A., 2002, MNRAS, 335, 487

\bibitem%[Maller, Dekel \& Somerville (2002)]{ref:mds02} 
Maller A.H., Dekel A., Somerville, R.S., 2002, MNRAS, 329, 423

\bibitem%[Moore et al (1999a)]{ref:Mo99a} 
Moore B., Ghigna S., Governato F., Lake G., Quinn T., Stadel J., Tozzi P., 
1999a, \apj, 524, L19 % substructure


\bibitem%[Moore et al. (1999b)]{ref:Mo99b} 
Moore B., Quinn T., Governato F., Stadel J., Lake, G., 1999b,
MNRAS, 310, 1147  % cusp -1.5

\bibitem%[Navarro, Frenk \& White (1996)]{ref:Nav96} 
Navarro J.F., Frenk C.S., White S.D.M, 1996, \apj, 462, 563

\bibitem%[Navarro, Frenk \& White (1997)]{ref:Nav97} 
Navarro J.F., Frenk C.S., White S.D.M, 1997, \apj, 490, 493

\bibitem%[Navarro \& Steinmetz (2000)]{ref:Nav00} 
Navarro J.F., Steinmetz M., 2000, ApJ, 538, 477

\bibitem%[Navarro et al. (2004)]{ref:Nav04} 
Navarro J.F., Hayashi E., Power C., Jenkins A., Frenk C.S., White S.D.M., 
Springel V., Stadel J., Quinn T.R., \mnras, submitted, astro-ph/0311231

\bibitem%[Power et al (2003)]{ref:Pow03} 
Power C., Navarro J.F., Jenkins A., Frenk C.S., White, S.D.M., Springel V.,
Stadel J., Quinn T.R., \mnras, 338, 14

\bibitem%[Press \& Schechter (1974)]{ref:Press74}
Press W.H., Schechter P., 1974, ApJ, 187, 425

\bibitem%[Schaap \& van de Weygaert (2000)]{ref:Schaap00} 
Schaap, W., van de Weygaert R., 2000, \aap, 363, L29

\bibitem%[Syer \& White (1998)] {ref:Syr98} 
Syer D., White S.D.M., 1998, \mnras, 293, 337

\bibitem%[Tanemura, Ogawa \& Ogita (1983)] {ref:Tan83} 
Tanemura M., Ogawa T., Ogita N., 1983, 
Journal of Computational Physics, 51, 191

\bibitem%[Taylor \& Navarro (2001)]{ref:Tay01}
Taylor J.E., Navarro J.F., 2001, ApJ, 563, 483

\bibitem%[Tremaine, H\'{e}non \& Lynden-Bell (1986)] {ref:Tre86}
Tremaine S., H\'{e}non M., Lynden-Bell D., 1986, \mnras, 219, 285

\bibitem%[van de Weygaert (1994)]{ref:Wey94} 
van de Weygaert R., 2000, \aap, 363, L29.

\bibitem%[Wambsganss et al (2004)]{ref:Wam04} 
Wambsganss J., Bode P., Ostriker J.P., \apj, sumbitted, astro-ph/0306088

\bibitem%[Weller et al (2004)]{ref:Wel04} 
Weller J., Bode P., Ostriker, J.P., in preperation.

\end{references}
\end{document}